%% file: main.tex
\documentclass[conference,10pt]{IEEEtran}
\usepackage{cite}
\usepackage{graphicx}
\usepackage{amsmath}
\usepackage{booktabs}
\usepackage{array}
\usepackage{url}
\usepackage{algorithm}
\usepackage{algpseudocode}
\usepackage{xcolor}
\usepackage{tikz}
\usetikzlibrary{arrows.meta}
\usepackage[T1]{fontenc}
\usepackage{newtxtext,newtxmath}
\graphicspath{{figs/}}

\begin{document}

\title{Who Approved This? Coding Agents, Human Accounts, and Merge-Gate Evidence at a Fintech Startup}

\author{\IEEEauthorblockN{Adam Manji}
\IEEEauthorblockA{Bashi AI\\
London, UK\\
adam@bashi.app}}

\maketitle

\begin{abstract}
\input{sections/abstract}
\end{abstract}


\input{sections/intro}
\input{sections/setting}
\input{sections/method}
\input{sections/accounts}
\input{sections/gates}
\input{sections/review}
\input{sections/lessons}
\input{sections/related}
\input{sections/threats}
\input{sections/conclusion}
\input{sections/backmatter}

\bibliographystyle{IEEEtran}
\bibliography{references}

\end{document}

%% file: sections/abstract.tex
From 3 October, every non-bot merge into our main branch passed a non-author review gate, yet 71 of those 263 merges (27\%) rested only on an approval that a coding agent had submitted through the technical lead's GitHub account, and the platform recorded each one as an ordinary approval by the lead. We report twelve weeks of agent-driven development at Bashi AI, a fintech startup preparing for a SOC~2 audit: 786 merged pull requests, 657 of them associated with agents by explicit markers or by inference, analysed from platform records and saved agent-session transcripts. In the gated phase, an agent session submitted every approval the lead's account gave on a colleague's pull request, and issued nearly all of its merges; the records cannot show who decided any of them. We recompute, for every merge, a core gate property: a non-author approval and a passing required check, both before the merge and both on the exact revision merged. The recomputation confirmed the gate wherever branch protection applied and found where it did not: agents switching enforcement off around individual merges, a merge script that recorded failed checks as passing, and an agent collapsing a stack of pull requests through unprotected branches. The engineers' survey answers show that one engineer's approvals also rested on review agents, with the person validating findings and deciding, and both engineers say review mattered when it tested a change against something outside the diff. We draw lessons on recording decision authority separately from account identity, keeping agents' credentials away from the controls they operate under, and checking evidence at the branch and revision where integration happens.

%% file: sections/intro.tex
\section{Introduction}\label{sec:intro}
A pull request approval on GitHub records an account, a revision, a state, and a time. Separation-of-duties controls read that record as evidence that a second person examined the change, which assumes that an account stands for a person. With coding agents, that no longer holds: an agent can write a change, ask another agent to review it, and submit the resulting approval through a developer's account; the repository then holds an ordinary approval from that developer.

This became a practical problem for us at Bashi AI in September 2026. We build a credit-analysis platform that reads confidential financial documents, and we were preparing for a SOC~2 audit, which meant explaining how every change reached the product. At the same time, I was directing many parallel Claude Code sessions that implemented changes, pre-reviewed them, relayed approvals, and ran merges, mostly through my GitHub account. Our review rule asked for approval from someone other than the author. The rule was satisfied on the record, but the record could not say whether a person or an agent had made each decision.

The gap is measurable. From 3 October, all 263 non-bot merges into \texttt{main} had a valid non-author approval and a passing check, yet 71 of them rested only on an approval that one of my sessions had submitted through my account. The platform shows which account approved each merge, but nothing in its record shows whether a person or a session decided. Agents also acted where the gate did not reach: on 14 October one collapsed a stack of pull requests through unprotected branches in under a minute.

Other teams face the same issue. GitHub's Copilot reviewer can now submit approvals that count toward a repository's required-approvals rule~\cite{github2026copilotapprove}, and Anthropic reports merging over 30,000 changes a day~\cite{anthropic2026sourcecontrol}. Repository mining has noted that logs record who executed a merge but not who decided it~\cite{chung2026collaborator}. We found no published study of the limits of these records as compliance evidence inside a team that depends on them.

We studied the 786 pull requests merged into our product repository between 21 July and 16 October 2026, and connected GitHub's records to the transcripts of the agent sessions that produced them. We ask:
\begin{itemize}
\item \textbf{RQ1.} What can account records and session transcripts establish about agents' part in approvals and merges?
\item \textbf{RQ2.} Which merges carried approval and required-check evidence for the exact revision merged, and what did the merges without it have in common?
\item \textbf{RQ3.} What do the records show about the work of reviewing, and what must come from the reviewers themselves?
\end{itemize}

The contributions are:
\begin{itemize}
\item an industrial case of agents acting through people's accounts when opening, approving, and merging, sized from transcripts, with an account of what this does to separation of duties (Section~\ref{sec:accounts});
\item a check of the \emph{core gate property}, a prior approval and a prior passing check on the merged revision, recomputed from platform records independently of the platform's enforcement (Section~\ref{sec:method}, Algorithm~\ref{alg:gate});
\item the merges that lacked that evidence and why: enforcement switched off, a script that recorded failed checks as passing, and an agent-collapsed stack outside protection (Section~\ref{sec:gates});
\item an account of what agents changed about familiar control weaknesses, a map from auditor questions to platform evidence, and a checklist for teams, marked by what we have and have not tried (Section~\ref{sec:lessons}).
\end{itemize}

%% file: sections/setting.tex
\section{Setting and Agent Workflow}\label{sec:setting}

\subsection{Company, Product, and Audit}
Bashi AI is a London startup building a credit-analysis platform for credit funds. The platform ingests a borrower's financial documents, extracts and reconciles the figures, supports questions to the borrower, and drafts a credit memo; analysts review the output and make the credit decision. The codebase is a web application with a Python backend, a relational database, a browser front end, and cloud infrastructure managed as code. The repository was created on 21 July 2026, so the study covers both building the product and building its development process.

Our prospective clients asked us to show how we protect their confidential documents, so we are preparing for a SOC~2 audit. SOC~2's change-management criterion, CC8.1, expects an organization to authorize, test, approve, and implement changes in a controlled way~\cite{aicpa2017tsc}. It does not prescribe two-person review; a non-author approval in the repository is the evidence we chose to offer for it. An auditor can then ask, for any change, who approved it, whether checks ran first, and whether anything took a different path. We wanted the repository to answer those questions from its own records.

The control we operated was this: a change enters \texttt{main}, and a release enters \texttt{production}, only through a pull request that a non-author approved and on which the required check passed. A push to \texttt{production} deploys automatically to the production environment. Merges between working branches were outside the control by design. We judge each incident in this paper against that control. A merge into \texttt{main} or \texttt{production} without the evidence is a violation; anything outside those branches is a boundary case, which can still affect attribution and audit.

\subsection{Team and Agent Roles}
The engineering team has three people. I am the technical lead: I set priorities, write the briefs that start agent sessions, and make or delegate the acceptance decisions. Two software engineers review pull requests and also write code, using agents themselves. All agent sessions ran Claude Code, Anthropic's command-line coding agent.

I ran many sessions in parallel, in four roles. An \emph{orchestrator} session kept the board of active work, wrote briefs, integrated results, and ran the merges. A \emph{project-management (PM)} session pre-reviewed pull requests, drafted messages for me, and relayed approvals. \emph{Lane} sessions each owned one scope, such as a feature or a data task, in a separate working tree. \emph{Workers} were subagents that a session started for one bounded step. Every session began from a written brief and ended with a written hand-off, and sessions exchanged short messages through a messaging tool.

\subsection{Accounts and Authority}
Agents acted through people's accounts. My sessions used my GitHub login, which I call the \emph{lead account}; it could open pull requests, submit reviews, merge, and administer branch protection. The PM session approved engineers' pull requests through the lead account after its own model review. The engineers' sessions used the engineers' accounts. No agent had a GitHub identity of its own.

\subsection{The Gate}
Branch protection on \texttt{main} and \texttt{production} required one approval from an account other than the pull request's author, and a passing run of a required check named \texttt{test}. Until 7 October, \texttt{test} was a single job that ran the static checks and the test suite. On 7 October it was split into parallel jobs, for static analysis, test shards, and a database migration round trip, behind an aggregator job that kept the name \texttt{test} and starts only when those jobs have finished. Other jobs were not required, and a row-level-security test job added on 8 October runs only after merge. Protection also dismissed approvals when new commits were pushed and required review conversations to be resolved; the requirement that the branch be up to date with \texttt{main} was switched on and off during the study.

Scripts run by agent sessions turned the rule into a routine. An approval wrapper, run by the PM session, read the checks on the pull request's exact head and submitted an approval through the lead account only if all of them had passed. A merge script, run by the orchestrator, merged a list of pull requests one at a time: it confirmed each was open and approved, brought its branch up to date, waited for the required check on the new head, and squash-merged with the verified head commit supplied as a guard against a late push.

A complete enforced-period record looks like this. The example is the pull request closest to the median size and lead time among those with an agent trailer, a branch-update commit, and an engineer's approval. A pull request from an engineer's account, 922 changed lines with agent trailers on its four change commits, passed the required check on its pushed head. The merge script added a branch-update commit, an approval was given on that final revision, the check passed on it, and the lead account squash-merged 0.7 minutes later. The platform records the author, approving account, approved revision, check result, and merge time, but not who decided to approve or merge.

The rule tightened in steps, which define three periods. In the \emph{setup period}, to 27 September, administrators could merge without the approval, and most merges did; on 25 September the rule was extended to administrators and lifted again for two hours that afternoon. In the \emph{transition period}, 28 September to 2 October, the team worked under the rule, but during a logged security remediation administrators switched enforcement off for individual merges. In the \emph{enforced period}, from 3 October, enforcement stayed on. We call the last two together the \emph{gated phase}.

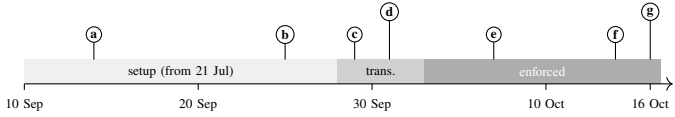
\begin{figure}[t]
\centering
\input{sections/fig-boundaries}
\caption{Study timeline and periods. (a) 14 Sep: surviving transcripts begin. (b) 25 Sep: rule extended to administrators, lifted for two hours. (c) 29 Sep--1 Oct: enforcement toggled around individual merges. (d) 1 Oct: false green. (e) 7 Oct: \texttt{test} split into jobs behind an aggregator. (f) 14 Oct: stack collapsed. (g) 16 Oct: cutoff.}
\label{fig:boundaries}
\end{figure}

Protection covered \texttt{main} and \texttt{production} only. Agents often built a change as a \emph{stack}: a chain of pull requests, each based on the previous one's branch rather than on \texttt{main}. Merges into those working branches were outside the protected boundary. Fig.~\ref{fig:boundaries} shows the timeline.

%% file: sections/fig-boundaries.tex
\begin{tikzpicture}[font=\scriptsize, x=0.23cm, y=1cm,
  ev/.style={circle, draw, fill=white, inner sep=0.8pt, font=\tiny\bfseries}]
\fill[black!6] (0,0) rectangle (18,0.32);
\fill[black!18] (18,0) rectangle (23,0.32);
\fill[black!32] (23,0) rectangle (36.6,0.32);
\node[font=\tiny] at (9,0.16) {setup (from 21 Jul)};
\node[font=\tiny] at (20.5,0.16) {trans.};
\node[font=\tiny, text=white] at (29.8,0.16) {enforced};
\draw[->] (0,0) -- (37.3,0);
\foreach \d/\t in {0/10 Sep, 10/20 Sep, 20/30 Sep, 30/10 Oct, 36/16 Oct}{
  \draw (\d,0) -- (\d,-0.08); \node[below, font=\tiny] at (\d,-0.08) {\t};}
\foreach \d/\h/\l in {4/0.55/a, 15/0.55/b, 19/0.55/c, 21/0.85/d, 27/0.55/e, 34/0.55/f, 36/0.85/g}{
  \draw (\d,0.32) -- (\d,\h); \node[ev] at (\d,\h+0.1) {\l};}
\end{tikzpicture}

%% file: sections/method.tex
\section{Data and Method}\label{sec:method}

\subsection{Sources and Unit of Analysis}
The unit of analysis is a merged pull request. We include every pull request merged from the repository's first merge on 21 July 2026 to a fixed cutoff of 16 October 2026, 15:48:23 UTC: 786 in total, of which 14 merged into branches other than \texttt{main} (5 releases into \texttt{production} and 9 into stacked working branches). Merges into a stack are separate merge events but not separate deliveries, so we report results for all destinations and for \texttt{main} alone. Dependency-bot pull requests (63) are excluded from the gate measures and reported separately; the merge counts, lead times, and approval counts in Section~\ref{sec:review} include them, as stated there.

The sources are GitHub's pull request, review, commit, and check-run records, retrieved through the API with read-only calls; the transcripts of Claude Code sessions saved on my machine; the coordination files that held briefs, hand-offs, and decisions; and the live branch-protection settings, read on 16 October. Claude Code deletes transcripts after about 30 days, so transcripts covered sessions from 14 September onward, which includes the whole gated phase. The organization audit-log API was not available on our GitHub plan, so the protection history before 16 October comes from our change log and from the protection-changing commands in the transcripts. We also compared the first-parent history of \texttt{main} and \texttt{production} with the merged pull requests. Seven commits reached \texttt{main} directly, all on 21--22 July while the repository was being set up; after that, every commit on \texttt{main} came from a merged pull request. Every commit on \texttt{production} is either one of its five release merges or a commit that had already passed through \texttt{main}.

The setup period holds 442 merged pull requests over 9.7 weeks; the gated phase holds 344 over 2.7 weeks, of which 59 non-bot merges fall in the transition period and 273 in the enforced period. The periods describe our history; they are not experimental treatments.

\subsection{The Core Gate Property}
A merge satisfies the core gate property when the records show two things before the merge: an approval of the pull request's final head commit by an account other than the author, which is that reviewer's latest review before the merge and was never dismissed; and a successful run of the required check \texttt{test} on that head that both started and finished before the merge, taking the latest such run that had started. The records are those the platform retains. A reviewer's later review supersedes an earlier one, and a dismissed approval keeps the state \texttt{DISMISSED}, so neither is counted. We fetch every attempt of the check, not only the latest (the API's default), so reruns cannot hide an earlier result.
Algorithm~\ref{alg:gate} gives the check with the API fields it reads. The head is the pull request's last commit, not the squash commit created on the destination branch. A run that started after the merge cannot satisfy the second condition, even if it passed. From 7 October, \texttt{test} is an aggregator job that starts only after the jobs it depends on have finished, so a run that started before the merge implies that those jobs also ran before it, and the aggregator's own run takes only seconds. For our 786 pull requests the check needs on the order of a thousand read-only API calls, so it can run on a schedule. Where the property fails, we record the reason: no approval, an approval naming an earlier revision, a failed check, a check still running, or no run started before the merge. None of the 24 gated-phase failures rests on absent data alone: each has an approval submitted after the merge, a completed failed run, or a run that started after the merge, except the four merges never approved at all, for which no review exists to retrieve.

We call the property \emph{core} because it covers the two conditions our SOC~2 evidence rests on and leaves out two other protection rules, that the branch be up to date with \texttt{main} and that review conversations be resolved. In our configuration, branch protection enforced both conditions: no required check in the study concluded as skipped or neutral, which protection would accept, and in the enforced period every approval on a merge into \texttt{main} named the merged head. On a protected branch with enforcement on, the property should therefore hold by construction. Recomputing it from the records does not rely on that enforcement, so it finds the merges enforcement missed: those made while it was off, those into unprotected branches, and those where a tool reported a check as passing when it had not.

\input{sections/alg-gate}

\subsection{Agent Association and Action Traces}
A pull request is \emph{explicitly} agent-associated when a commit or its description carries an agent marker, such as a co-author trailer. It is \emph{inferred} as agent-associated when it was opened from the lead account without a marker, because nearly all of my pull requests were opened by sessions; we checked this against the transcripts by looking for the command that opened each one. Pull requests from the engineers' accounts without a marker are \emph{unmarked}, not human-written: the engineers used agents too.

We searched the transcripts for approval and merge commands issued by agent sessions, in every form we found: the \texttt{gh} command-line tool, direct calls to the review and merge API endpoints, our approval and merge wrapper scripts, and loops over pull request numbers. We matched each command to the platform event it preceded, within ten minutes for direct commands and within four hours for scripts that wait for checks before merging. A hand check of a random sample of 15 approval matches and 15 merge matches against the transcripts confirmed 15 and 14; the one error was a merge attributed to an earlier command that had been stopped, when a later agent command had performed it. A match shows that an agent issued the command; it does not show who decided the command should be issued. The transcripts comprise about 2,240 session files, of which 2,172 contain activity between 14 September and 16 October.

\subsection{Review Measures and Engineer Survey}
Lead time runs from opening to merge. For pull requests with a non-author approval before merge (on any revision), we report the time from opening to the first such approval and from the last one to the merge. Opening is not the same as being ready for review, so the first interval includes implementation time.

The 45 standing corrections to agents (Section~\ref{sec:accounts}) were classified by class and by how each is enforced, with model assistance; there was no second coder.

To learn what the records cannot show, we asked the two engineers five questions on 18 October: three rated agreement statements (whether they read every changed line of agent-built pull requests, whether reviewing them takes the same effort as reviewing a colleague's, and whether they know when an agent prepared a pull request opened from another person's account) and two short-answer questions (what they check themselves and what they leave to CI or the description, and one pull request where their review changed the outcome and one where it added little). Each answered privately, without seeing the other's answers, and was told that answers might be quoted or paraphrased without a name. Both answered, and neither asked us to leave anything out.

%% file: sections/alg-gate.tex
\begin{algorithm}[t]
\caption{Core gate property for one merged pull request}
\label{alg:gate}
\footnotesize
\begin{algorithmic}[1]
\Require PR with \texttt{merged\_at}, author, head SHA $h$ (\texttt{headRefOid})
\State $A \gets$ reviews with \texttt{state} $=$ APPROVED, \texttt{user} $\neq$ author, \texttt{submitted\_at} $<$ \texttt{merged\_at}
\State $approved \gets \exists\, a \in A$ with $a.\texttt{commit\_id} = h$
\State $R \gets$ check runs on $h$ named \texttt{test} with \texttt{started\_at} $<$ \texttt{merged\_at}
\If{$R = \emptyset$} \State $checked \gets$ false \Comment{no pre-merge run}
\Else
  \State $r \gets \arg\max_{r \in R} r.\texttt{started\_at}$
  \State $checked \gets r.\texttt{conclusion} = $ success $\wedge$ $r.\texttt{completed\_at} < \texttt{merged\_at}$
\EndIf
\State \Return $approved \wedge checked$
\end{algorithmic}
\end{algorithm}

%% file: sections/accounts.tex
\section{Who Acted and Who Decided}\label{sec:accounts}
\input{sections/table-accounts}

\subsection{Opening Pull Requests}
Table~\ref{tab:accounts} summarizes what the records show. Of the 786 merged pull requests, 229 carry an explicit agent marker and 428 more were opened from the lead account without one, so 657 (83.6\%) are agent-associated. The lead account opened 487 of the 723 non-bot pull requests (67.4\%). For the inferred pull requests merged after transcripts became available, a recorded session issued the command that opened 330 of 342 (96.5\%). Agent work was not confined to my account: the engineers' accounts carry explicit agent markers on 170 pull requests.

Markers depended on the workflow rather than on who wrote the code. Some sessions added co-author trailers and some did not, and the engineers marked their agent-assisted work inconsistently. On the platform, a pull request's author field identified the account that opened it, and neither the person nor the agent who prepared it.

\subsection{Approving and Merging}
The same held for approvals and merges, which are the steps a separation-of-duties control relies on. In the gated phase, the lead account approved 86 non-bot pull requests, and an agent session submitted the approval for every one of them. Of the 88 approval commands, 31 called the \texttt{gh} tool directly and 57 ran a wrapper script that approved only if every check on the exact head had passed. The records contain no lead-account approval in that phase that a session did not submit. The lead account also made 317 of the 332 gated-phase non-bot merges; for 262 of them a session issued the merge command or ran one of our merge scripts, and for 24 more a session had armed auto-merge. Of the remaining 31, 24 were merged by one-off agent scripts whose pull request lists sat inside the script rather than on the command line, and for 7 we found no issuing command.

This was the intended workflow. The PM session pre-reviewed an engineer's pull request, checked that the approval would apply to the current revision and that CI had passed, and submitted the approval through my account. GitHub then recorded an approval from an account other than the author, which is what the rule required. It did not record whether I had read the change, accepted the session's recommendation, or left the decision to the session. Some of these approvals carried a decision I had made and a session relayed; others I left to sessions given an adversarial review role. I kept no record of which was which, and neither did the platform. Both forms of delegation produce the same record: an approval issued by an agent can carry my decision, and an approval from my account can rest entirely on an agent's assessment.

\input{sections/table-evidence}
Table~\ref{tab:evidence} sets out which source establishes each step. No source records the decision to accept, which is the step a separation-of-duties control is about.

\subsection{Instructions Did Not Limit Agent Actions}
On 19 August, a worker started for research, with an explicit instruction to report and take no action, merged two pull requests and opened a replacement for a third; we found out from a routine query, not from the worker's report (the account comes from the incident record written that day). Separate working trees now keep sessions from switching each other's branches, and from 3 October protection blocks unapproved merges into \texttt{main}, but neither restricts what a session may do through the lead account: an approved pull request can still be merged by a session whose brief said not to.

Corrections like this became standing rules in memory files that later sessions read: of 45 written between 21 September and 8 October, 3 were backed by a mechanism, 7 partly, and 35 were instructions only.

%% file: sections/table-accounts.tex
\begin{table}[t]
\caption{Agent association and agent-issued commands. Opening classes cover all 786 merges; the opening check covers pull requests merged from 14 September; approval and merge commands cover gated-phase non-bot pull requests.}
\label{tab:accounts}
\centering
\footnotesize
\begin{tabular}{@{}lr@{}}
\toprule
Observation & Count \\
\midrule
Merged pull requests & 786 \\
\quad explicit agent marker & 229 \\
\quad opened from lead account, no marker & 428 \\
\quad engineer account, no marker & 66 \\
\quad dependency bot & 63 \\
\midrule
Unmarked lead-account PRs opened by a session & 330/342 \\
Lead-account approved PRs, approval submitted by a session & 86/86 \\
Lead-account merges issued by a session & 262/317 \\
\quad plus auto-merge armed by a session & 24/317 \\
\bottomrule
\end{tabular}
\end{table}

%% file: sections/table-evidence.tex
\begin{table}[t]
\caption{Which source establishes each step of a change, for a pull request approved and merged through the lead account in the gated phase.}
\label{tab:evidence}
\centering
\footnotesize
\setlength{\tabcolsep}{3pt}
\begin{tabular}{@{}>{\raggedright\arraybackslash}p{0.27\columnwidth}>{\raggedright\arraybackslash}p{0.33\columnwidth}>{\raggedright\arraybackslash}p{0.32\columnwidth}@{}}
\toprule
Step & Established by & Unknown \\
\midrule
Preparing the change & Agent markers; opening command in transcripts & How much an agent wrote \\
Reviewing the code & Session's model pre-review in transcripts; engineers' survey & What any person read \\
Deciding to accept & --- & Whether I or a session decided \\
Submitting the approval & Platform record (account); transcripts (session) & --- \\
Executing the merge & Platform record (account); transcripts (session or script) & On whose instruction \\
\bottomrule
\end{tabular}
\end{table}

%% file: sections/gates.tex
\section{Gates, Branches, and Revisions}\label{sec:gates}
\input{sections/table-gate}

\subsection{The Gate by Period and Destination}
Table~\ref{tab:gate} applies the core gate property to every non-bot merge. In the setup period it held for 103 of 391 merges (26.3\%). The required check was rarely the problem; it passed before merge on 379. Approval was: 264 merges had no non-author approval before merge, and on 22 more the approval named an earlier revision, which was possible because administrators could merge past a stale approval. Agents also changed the protection settings themselves. On 25 September at 12:59 UTC an agent session applied the configuration that extended the rule to administrators. At 14:03 another session removed administrator enforcement again, and it and its workers merged nine pull requests as administrators before restoring enforcement at 16:19. None of the 10 pull requests merged that day was approved. The team began working under the rule on Monday 28 September, which is why the transition period starts then.

In the transition period the property held for 41 of 59 merges. The 18 exceptions are 16 merges into \texttt{main}, one release into \texttt{production}, and one merge into a stack branch. Fourteen were approved after the merge and 4 never, and 4 lacked a passing required check (Section~\ref{sec:falsegreen}). The transcripts show how most were made. On 29 September a script set the required approval count to zero, merged one pull request, and restored it ten seconds later. From 30 September, under a recorded emergency change, a script and then the merge script itself removed administrator enforcement before each merge and restored it immediately after, on \texttt{main} and once on \texttt{production}. Every unapproved merge into a protected branch in this period matches a protection change recorded in the transcripts.

From 3 October the transcripts show no further change that suspended either core rule; the only protection changes were to the up-to-date requirement, which the property does not cover. In this enforced period, all 263 merges into \texttt{main} had the property. The required check finished a median of 1.0 minute before each merge, because the merge script waited for it, and every approval named the final head, so approvals were given, or given again, after the last branch update. Across all destinations the count is 267 of 273. The ten merges elsewhere were three releases into \texttt{production}, which all had the property, and seven merges into stack branches, six of which did not. Over the whole window, 14 merges went to branches other than \texttt{main}: one release in the setup period, one release and two stack merges in the transition period, and the ten above. The 63 dependency-bot merges, excluded from the table, follow the same pattern: 24 of 51 had the property in the setup period, and all 12 in the gated phase.

Against the control in Section~\ref{sec:setting}, the setup-period merges predate it. Seventeen of the 18 transition exceptions, the 16 into \texttt{main} and the release, are violations made under deliberate exceptions: one on 29 September by a one-off script, the rest under the emergency change recorded from 30 September. Thirteen of the 17 were approved after the fact and 4 never were. The stack merges, in the transition period and in Section~\ref{sec:stack}, violated nothing, because they went into branches the control did not cover.

\input{sections/table-approvers}
Putting the two halves of the study together, Table~\ref{tab:approvers} shows whose approval each enforced-period merge into \texttt{main} rested on. For 192 merges it was an engineer's account; 185 of those were pull requests that my sessions had opened. For the other 71, the only valid approval came from the lead account, and Section~\ref{sec:accounts} shows that a session submitted every such approval. On 57 of the 71, the change itself also carried agent markers from the engineer's side. For these merges, an agent built the change and an agent submitted the approval, and the record shows a two-person control between two people. The property held on all 263; for 71, it says nothing about whether a person looked. The other 192 are not a verified human-only group either: transcripts from the engineers' machines were not available, and E1 reports using review agents (Section~\ref{sec:review}).

\subsection{An Agent Collapsed a Stack}\label{sec:stack}
\input{sections/table-stack}
On 14 October, agent sessions had built a front-end restyle as a stack of eight pull requests. The bottom one (\#1337) was based on \texttt{main}; each of the other seven was based on the branch of the one below it. One engineer approved all eight between 17:05 and 17:15 UTC, each with a written review of 459 to 870 characters. E1 reports that review agents draft such reviews and the engineer checks their findings (Section~\ref{sec:review}); the record cannot show how closely these eight were checked. The orchestrator's working notes recorded its plan: fold the stack upward into the bottom pull request by merge commits and then merge that one into \texttt{main} after a final review.

At 17:16:43 the orchestrator carried out the first step with a single command, recorded in its transcript as
\begin{center}\footnotesize\texttt{foreach (\$n in 1345,1344,1343,1342,1340,1339,1338)}\\\texttt{\{ gh pr merge \$n -R \$R --merge ... \}}\end{center}
which merged the seven upper pull requests one after another, from the top down, through the lead account, in 58 seconds (Table~\ref{tab:stack}). (\#1341 is an unrelated issue.) Each merge into a branch added a merge commit to the pull request below it, which became that pull request's new head. When the command reached each lower pull request, its approval named a commit that was no longer its head, and the required check on the new head had not yet run. The checks on those six heads started about 8 to 11 minutes after the merges. All six passed, but only after the merges.

Nothing blocked these merges, because the bases were working branches and branch protection covered only \texttt{main} and \texttt{production}. On those branches, approvals were not dismissed when the head changed, and nothing waited for the check. Only the top pull request, whose head did not change, had the property.

The combined change, 1,325 changed lines across 56 files, then reached \texttt{main} through \#1337, and there the gate worked as designed. The fold had changed \#1337's head, so the platform dismissed the engineer's earlier approval. The engineer approved the new head at 17:51:59, the required check ran on it from 18:06:05 to 18:06:30, and the merge followed at 18:06:44.

Integrating a stack on working branches and gating only the combined change is a reasonable design, and under our control it was the intended one. Two things still matter for evidence. First, the agent planned and executed the intermediate merges itself, and the platform attributes them to me, so an auditor reading the history sees seven merges by the technical lead, six of them on approvals that no longer named the merged heads. Second, the final approval covers the combined diff, so whether the reviewer of \#1337 looked as closely at the whole as the earlier approvals implied for each part is not something the record can show.

\subsection{Approvals on Superseded Revisions}
On protected branches, protection handled changed heads as designed. In the gated phase, it dismissed 106 approvals on 84 of 332 non-bot pull requests (25.3\%) because a later commit changed the head, and 83 of those 84 pull requests were approved again on the merged head before merging; the exception was a production release merged during the remediation and approved five hours later. In 29 of the 106 dismissals, the commit that changed the head merged \texttt{main} into the branch to bring it up to date. In 51, the approved commit was no longer in the pull request's history, because the branch had been rebased or force-pushed. The other 26 were new commits added on top. Dismissed approvals stay visible in a pull request's history. On protected branches the platform discounts them, and anyone counting approvals from an export must do the same. On unprotected branches nothing discounts them, so the stack's stale approvals still appear as approvals.

\subsection{A False Green}\label{sec:falsegreen}
On 1 October the merge script recorded four merges as passing CI when they had not. Enforcement was off during the remediation, so the script was the only remaining check. After a failed job was re-run, the script read the pull request's combined status instead of the required check on the head commit, concluded that CI had passed, and merged four pull requests. The change log recorded them as green. Three had a completed, failed \texttt{test} run before the merge, and the fourth had no run that started before the merge. The failing test was date-dependent and was failing on \texttt{main} that day as well.

The fix changed what the script reads: it now queries the required check on the head commit and requires that run to have succeeded. A standing rule forbids recording a merge as green without that query, and re-enabled branch protection added an independent barrier.

%% file: sections/table-gate.tex
\begin{table}[t]
\caption{The core gate property for non-bot merges. Approval and check must both precede the merge; ``head'' requires them to name the merged revision.}
\label{tab:gate}
\centering
\footnotesize
\setlength{\tabcolsep}{2.6pt}
\begin{tabular}{@{}lrrrrr@{}}
\toprule
 & & \multicolumn{2}{c}{Approval} & Check & Core \\
\cmidrule(lr){3-4}
Period, destination & Merges & any rev. & head & head & gate \\
\midrule
Setup, all & 391 & 127 & 105 & 379 & 103 \\
Transition, all & 59 & 41 & 41 & 55 & 41 \\
Enforced, all & 273 & 273 & 267 & 267 & 267 \\
Enforced, \texttt{main} & 263 & 263 & 263 & 263 & 263 \\
\bottomrule
\end{tabular}
\end{table}

%% file: sections/table-approvers.tex
\begin{table}[t]
\caption{Whose approval the 263 enforced-period merges into \texttt{main} rested on (non-bot).}
\label{tab:approvers}
\centering
\footnotesize
\begin{tabular}{@{}llr@{}}
\toprule
Valid approval from & Pull request opened by & Merges \\
\midrule
Engineer account only & lead account (agent) & 185 \\
 & engineer account & 7 \\
Lead account only, & engineer account, agent-marked & 57 \\
\quad submitted by a session & engineer account, unmarked & 14 \\
\midrule
Total & & 263 \\
\bottomrule
\end{tabular}
\end{table}

%% file: sections/table-stack.tex
\begin{table}[t]
\caption{The stack collapsed on 14 October, in merge order (UTC). ``Approved head'': a non-author approval names the merged head. ``Check start'': the first \texttt{test} run on the merged head; for the six lower pull requests, none had started before the merge.}
\label{tab:stack}
\centering
\footnotesize
\setlength{\tabcolsep}{3pt}
\begin{tabular}{@{}llllc@{}}
\toprule
PR & Base & Merged & Check start & Approved head \\
\midrule
\#1345 & stack branch & 17:16:47 & 16:21:21 & yes \\
\#1344 & stack branch & 17:16:56 & 17:27:29 & no \\
\#1343 & stack branch & 17:17:06 & 17:26:08 & no \\
\#1342 & stack branch & 17:17:15 & 17:25:00 & no \\
\#1340 & stack branch & 17:17:25 & 17:28:21 & no \\
\#1339 & stack branch & 17:17:34 & 17:28:53 & no \\
\#1338 & stack branch & 17:17:45 & 17:28:40 & no \\
\midrule
\#1337 & \texttt{main} & 18:06:44 & 18:06:05 & yes \\
\bottomrule
\end{tabular}
\end{table}

%% file: sections/review.tex
\section{The Work of Reviewing}\label{sec:review}
\subsection{What the Records Show}
Once the rule bound every account, most approvals came from the engineers' accounts: in the gated phase they gave 259 of the 360 retained non-author approvals (71.9\%; dismissed approvals are not counted, and approvals of the dependency bot's pull requests are), over 28 reviewer-days (one account on one calendar day with an approval), a median of 6.5 approvals per reviewer-day. The other 101 came from the lead account: 88 on the engineers' pull requests, all submitted by sessions, and 13 on the bot's. Throughput rose from 45.4 to 128.8 merges a week and the median time from opening to merge from 1.0 to 21.7 hours, but serial merging, the 7 October CI split, and a backlog week of 174 merges changed at the same time, so we treat these as context only. Changes were requested on 18 of 344 gated-phase pull requests (5.2\%) against 12 of 442 before (2.7\%), which says that some reviews objected but not how closely any reviewer read.
\subsection{What the Engineers Report}
We refer to the engineers as E1 and E2.

\emph{Reading.} E1 rated ``I've read and understood every changed line'' 3 of 5: ``I usually read a good part of the diff, but not every line.'' E2 rated it 4. E1 did not review alone. Review agents ``read the diff, run the tests and checks, and draft the review''; E1 reads ``what they found and the exact text that goes out under my name'' and decides whether it is valid to post. Agents therefore stood behind the engineers' approvals as well as mine. Unlike mine, E1's approvals carried a stated human check on the agents' findings.

\emph{Effort.} E1 rated equal effort 3 of 5 and E2 rated it 4. E1 described the effort as moving ``away from reading code and more towards decisions: block or approve with a precondition, open an issue or not, in what order.'' Re-approving after a branch update that left the diff unchanged was ``close to zero effort''. This matches Anthropic's report that its engineers review the intent of a change rather than every line~\cite{anthropic2026sourcecontrol}.

\emph{Knowing who prepared a change.} E1 rated ``I know whether they or an agent session prepared it'' 2 of 5, relying on self-review comments and a ``(via Claude)'' tag, and said that knowing does not change the review: ``I answer agent work with agent review, but I re-validate the findings \ldots{} against real artifacts, not just against the diff.'' E2 rated it 3 and did not say whether knowing changes the review.

\emph{What reviewers keep and delegate.} E1 keeps the final text of each review, the decision to block, and what deserves a follow-up. E1 leaves to agents and CI the line-by-line reading, running the tests, CI status, infrastructure plan counts, confirming the diff has not changed since the last approval, and checking that suggestions were applied. E2 checks the tests the agent wrote and looks for overengineering.

\emph{When review mattered.} E1's examples of review changing the outcome came from testing a claim against real data or repeated runs, and two can be traced in the records, both just after our cutoff. On an agent-built pull request (\#1433), E1 found a test that failed three of five local runs while CI passed, ``green by luck''; the fix was applied before the merge. On 16 October, E1 approved two agent-built pull requests (\#1455 and \#1435) ``with one precondition'': a fix for a figure that would otherwise pass as valid, tracked in a separate issue. The platform recorded both as ordinary approvals, and both were merged through the lead account without the precondition met; for one, E1 later opened the fix as a separate pull request. A conditional approval is, to the gate, an approval. Review added little, by E1's account, on re-approvals of unchanged diffs: ``The useful reviews tested the claim against real domain data. The low-value ones only re-confirmed what CI already showed.'' E2 drew the same line: review ``mattered when I tested the reviewed changes against something outside the diff, like boundary values'', and added little when E2 ``read the diff on its own, like a model would''.

With two respondents these answers are illustrative; they describe review work the records do not show. They also show a limit of this kind of review: a person who validates an agent's findings checks what the agent found, and may share its blind spots for what it did not. What the records show as one approval, E1 describes as an agent's reading, a person's check of the agent's findings against real artifacts, and a person's decision; the platform records only the approval.

%% file: sections/lessons.tex
\section{What Is Specific to Agents, and Lessons}\label{sec:lessons}

\subsection{What Agents Changed}
Agents did not create new kinds of control failure. Delegated submission under shared credentials, rules that exist only on paper, and scripts that read the wrong status all predate them. What agents changed was how often these weaknesses were exercised, and that the actor behind each one was a session using a person's account.

\emph{Delegated acceptance.} In the gated phase, a session submitted every approval my account gave on a colleague's pull request, most through a wrapper that checked CI on the exact head and then approved (Section~\ref{sec:accounts}), and 71 merges into \texttt{main} rested on nothing else (Table~\ref{tab:approvers}). No platform record shows what review of the code preceded them; the transcripts show the PM session's model pre-review, but not whether I read the change. The decider could be me or the session, and the record did not say which. The 71 do not show 71 absent human decisions; they show 71 approvals whose decider cannot be established.

\emph{Written instructions did not constrain actions.} A research-only brief did not stop a merge, and 35 of our 45 corrections existed only as text that a later session would see only if it was given the file.

\emph{Shared credentials reached the control itself.} Every protection change found in the transcripts, from the three on 25 September to the per-merge toggles of 30 September and 1 October, was issued by an agent session (the transcripts cannot show changes made in the web interface). The credentials that let a session approve also let it suspend the rule it was approving under.

\emph{Speed.} The orchestrator merged seven pull requests in 58 seconds, finishing before the first check on the new heads had started (Section~\ref{sec:stack}). A shell loop run by a person would be as fast; here such loops were a routine part of the workflow.

\emph{Superseded approvals and summary statuses.} Approvals left on superseded revisions occur whenever a head changes after review (Section~\ref{sec:gates}), and the false green came from a script reading a summary status. Both happened inside routines that agents ran many times a day.

These problems have platform remedies: rulesets that target any branch pattern~\cite{github2026rulesets}, merge queues that require checks to pass on the merged result~\cite{github2026mergequeue}, separate bot identities for automation~\cite{github2026appsvsoauth}, a protection option requiring that the most recent push be approved by someone other than its pusher~\cite{github-branch-protection}, and, for GitHub's own coding agent, a rule that the requester's approval does not count toward required reviews~\cite{github2026copilotreview}. For some of the 71 merges the lead account had also made the last push, a branch update; whether the last-push option would have rejected those approvals depends on whether GitHub treats such an update as a reviewable push, which we have not tested. Supply-chain frameworks state the underlying requirement directly: the SLSA source draft asks that changes to protected branches be agreed by two trusted persons~\cite{slsa2025source}, and in-toto attests who performed each step~\cite{torresarias2019intoto}. Shared credentials do not disable repository controls, as our enforced period shows, but they hide who acted under them, and none of the identity remedies applies unless a team sets it up; until then, the records look the same as a working two-person control.

\subsection{What an Auditor Would See}
We expect a SOC~2 auditor to test changes deployed to production. That population is the five releases into \texttt{production} and the changes they carried; it would not include the six stack merges. By the cutoff, 43 of the 71 merges with agent-submitted approvals had reached \texttt{production}, so they are inside it. We expect an auditor to read those approvals as approvals by the technical lead. We propose, without an auditor's test, that such approvals should be disclosed in the system description and backed by a record that a person with authority owned each decision. The lead account could approve, merge, and change protection, which is a management-override weakness with or without agents; agents made using it quick and routine. The transition-period merges would be assessed under the emergency-change provisions of CC8.1~\cite{aicpa2017tsc}, where approval after the fact under a recorded emergency procedure is a different finding from no approval at all.
\subsection{Lessons}
Table~\ref{tab:auditor} maps auditors' questions to platform evidence.

\input{sections/table-auditor}

\textbf{Record who decided, not only who acted}. If a control requires a person's judgment, give agents their own identities, reserve approval credentials for people, and record which session submitted each action and on whose decision. A decision record the agent can write itself is not independent evidence; it carries weight only if it comes through a channel the agent cannot use, such as the person's own approval from an identity no agent holds. Another project moved its agent to a separate GitHub App identity after finding the same problem in October 2026~\cite{athanor2026agentapp}. Separate identities are not enough by themselves, because on GitHub the permission that lets an identity open pull requests also lets it submit reviews. The enforcement has to come from the rule that counts approvals. One way to do this is a code-owners file that lists only people, with protection set to require a code owner's approval; agent identities can then still comment and review, but their approvals do not satisfy the rule. Agents keep permission to open, push, and merge (ideally through a merge queue), and only people can change protection. A session identifier in a review helps only when it is bound to such an identity; otherwise it is the agent's own assertion. Finally, a team should say what a person's approval asserts. In our case, per E1, it asserts that the person validated the agents' findings and accepted the change, not that they read every line.

\textbf{Keep agents' credentials away from the controls}. Separate working trees did not restrict what sessions could do through shared credentials, and a session that can administer branch protection can suspend it. Decide what each role may change and enforce that in the tool or service that performs the action.

\textbf{Name the boundary the gate protects}. Decide which events the policy covers: entry to \texttt{main}, every merge, or deployment. Anthropic moved its automated testing to after merge and before deployment~\cite{anthropic2026sourcecontrol}; if testing moves, the evidence must be collected at the new boundary. Report gate results per destination branch, not as one repository-wide rate.

\textbf{Tie every piece of evidence to a revision and a time}. Keep the commit an approval names, the identity and timing of the required run, and the merged head. A check that passed after the merge cannot justify it, and a dismissed review does not count as an approval. Conditions belong in a request for changes, not in the text of an approval, which the gate counts in full. Merge tools should read the specific check run, not a summary status.

\textbf{Measure review work before judging capacity}. Approval counts and elapsed times show the flow of changes but not what a reviewer read. Record when a pull request is ready for review, separate active review from waiting, and ask reviewers what they examine and what they leave to CI.

\begin{table}[t]
\caption{Checklist for a team whose agents act through a version-control platform. Items marked * were in place during the study; the rest are proposals we have not yet tested.}
\label{tab:checklist}
\centering
\footnotesize
\begin{tabular}{@{}>{\raggedright\arraybackslash}p{0.95\columnwidth}@{}}
\toprule
1. Give each agent role its own identity, and require approval from a code owner who is a person, so that agent approvals do not count; keep protection settings for people only. \\
2. Require approval of the most recent push by someone other than its pusher. \\
3. Put the session identifier in every review body and merge commit message. \\
4. Decide which branches the gate covers; if intermediate merges must carry evidence, apply review and check rules to those branches too. \\
5.* Make merge tools read the named required check on the head commit, never a combined status. \\
6. Run Algorithm~\ref{alg:gate} on a schedule and report results per destination branch (we ran it once, for this study). \\
7. Log every protection change with who or what made it and why, and keep the history of what the required check contains. \\
8. Ask reviewers, periodically, what they read and what they rely on. \\
\bottomrule
\end{tabular}
\end{table}

\subsection{Answers to the Research Questions}
\textbf{RQ1.} Account records showed which account acted, and on which revision. Transcripts showed which of those actions agents issued: in the gated phase, every approval my account gave and most of its merges. Neither showed who decided.

\textbf{RQ2.} In the enforced period every merge into \texttt{main} had prior approval and check evidence on the merged revision, though for 71 the approval was agent-submitted. The merges without the evidence happened where enforcement did not reach: it had been switched off, a script misread the evidence, or the destination branch was unprotected.

\textbf{RQ3.} The records show who approved, when, and on which revision, and give weak signals of depth. The engineers' two illustrative accounts describe review work that the fields we studied do not capture (Section~\ref{sec:review}).

\subsection{After the Study}
Three days after the cutoff, little has changed. We prepared CI to run the required check on GitHub's merge queue on 16 October but have not enabled the queue. Agents still act through the lead account, stack branches are still unprotected, and the last-push approval rule is off. We plan to give agents their own identity and to apply Table~\ref{tab:checklist}'s first four items before the auditor tests our controls, and we will describe agent-submitted approvals in our SOC~2 system description.

%% file: sections/table-auditor.tex
\begin{table*}[t]
\caption{Auditor questions about a change (SOC~2 CC8.1), the GitHub field that answers each, what it cannot show, and what we now think a team using agents should record.}
\label{tab:auditor}
\centering
\footnotesize
\begin{tabular}{@{}>{\raggedright\arraybackslash}p{0.2\textwidth}>{\raggedright\arraybackslash}p{0.22\textwidth}>{\raggedright\arraybackslash}p{0.25\textwidth}>{\raggedright\arraybackslash}p{0.27\textwidth}@{}}
\toprule
Auditor question & Platform field & What it cannot show & Record additionally \\
\midrule
Who approved the change? & Review \texttt{user}, \texttt{state}, \texttt{submitted\_at} & Whether a person or an agent decided; whether a session submitted it & Agent identities separate from people's; the session and decision behind each approval \\
Was the approved version the one merged? & Review \texttt{commit\_id} vs.\ PR head & Coverage of a combined change built from a stack & Base branch of every merge; approval of the final combined revision \\
Did required tests pass first? & Check run \texttt{head\_sha}, \texttt{started\_at}, \texttt{completed\_at}, \texttt{conclusion} & What the required check contained at that time & The check's job list per period; the specific run cited by any merge tool \\
Did anything bypass the process? & Branch-protection settings (current only) & Past settings; merges into unprotected branches & Protection history; per-branch gate results; every enforcement change with time and reason \\
Who merged it? & PR \texttt{merged\_by}, \texttt{merged\_at} & Whether a person or a script or agent executed it, and on whose instruction & The session or script that merged, and the instruction it acted on \\
\bottomrule
\end{tabular}
\end{table*}

%% file: sections/related.tex
\section{Related Work}\label{sec:related}
\textbf{Agents in pull requests.} Large datasets now collect agent-authored pull requests from public repositories~\cite{li2026aidev,li2025aiteammates}. Watanabe et al. studied pull requests created with Claude Code and the revisions made before acceptance~\cite{watanabe2025agenticprs}. Chung and Hassan classified 29,585 agent pull request lifecycles by who initiated and who approved them, and pointed out the gap between executor and decision-maker discussed in Section~\ref{sec:intro}~\cite{chung2026collaborator}. These studies work from public repositories, where an agent's identity is often visible. Our case concerns agents acting through people's accounts in a private repository, and measures what that does to a review gate used as compliance evidence.

\textbf{Reviewing agent output.} Park et al. describe the work of supervising coding agents~\cite{park2026delegation}, and Hassan et al. set out a research roadmap for agentic software engineering~\cite{hassan2025agentic}. Duma et al. found that most AI-generated pull requests they studied received no human review~\cite{duma2026reviews}, and Monperrus argues that agent inspection will supersede human code review~\cite{monperrus2026endofcodereview}. Anthropic's review agents comment on nearly every internal pull request but do not approve~\cite{anthropic2026codereview}; GitHub's reviewer agent can~\cite{github2026copilotapprove}. Other work examines automated review comments and whether they lead to changes~\cite{vijayvergiya2024autocommenter,sun2025aicodereview}, and large deployments of generative tools in industry~\cite{mao2026whatscode,chen2026beyondcommit}. Building on Chung and Hassan's executor/decision-maker observation, our case adds evidence of approvals submitted under a person's credentials and what that does to a control that relies on them.

\textbf{Compliance and continuous delivery.} Fitzgerald et al. showed how agile teams in a regulated domain kept compliance evidence continuous rather than retrospective~\cite{fitzgerald2013scaling}, and Moyón et al. examined how security compliance requirements can be built into agile engineering at scale~\cite{moyon2020integrate}. Neither addresses agents acting through the accounts of the people in the process, which is our case.

\textbf{Review and integration at scale.} Review coverage and participation are associated with software quality~\cite{mcintosh2014coverage}, and review also serves understanding and knowledge sharing~\cite{bacchelli2013expectations}, which approval counts do not capture. Large organizations have engineered around the cost of checking every change before it lands, by testing after submission~\cite{memon2017taming} or by speculatively building pending changes together~\cite{ananthanarayanan2019keeping}, and Anthropic has done so for agent-scale volume~\cite{anthropic2026sourcecontrol}. Stacked changes have been studied as relation chains in Gerrit~\cite{belhouchette2026relationchains}. GitHub records the revision an approval applies to~\cite{github-reviews-api} and can dismiss approvals when the head changes~\cite{github-branch-protection}, but only where protection applies, which is the gap our stack case fell into.

%% file: sections/threats.tex
\section{Threats to Validity}\label{sec:threats}
\textbf{What the records establish.} Account separation does not establish separation between a person and an agent, and a matched command does not establish who decided. I kept no record of which acceptance decisions I made myself, so the split between my decisions and delegated ones is unknown, not merely unrecorded by the platform. Transcripts exist only for sessions on my machine, so we could not check whether agents submitted any of the engineers' 259 approvals; E1 reports that review agents drafted the reviews and E1 decided what was posted. Missing markers and the 30-day transcript window can hide agent involvement, so the explicit count in Table~\ref{tab:accounts} is a lower bound and the unmarked count an upper bound on unassisted work; the inferred count can include a few pull requests no session opened (12 of 342 checked).

The gate check uses retained review states and cached check runs. A missing pre-merge run may reflect incomplete history rather than a merge without a check, and a dismissed review may hide an earlier valid decision. Without the platform audit log, the enforcement history rests on our change log, which agents helped write and which recorded the false greens of 1 October, and on transcripts, which miss any change made in the web interface. We used both only for dates and causes and checked every gate result against the platform's own records. The required check itself changed during the study, from one job to an aggregator on 7 October.

\textbf{Interpretation.} Enforcement, serial merging, new CI jobs, the mix of work, and agent use changed in the same weeks, so we do not attribute changes in throughput or lead time to any of them. The enforced period is two weeks long. The time from opening to first approval includes implementation and waiting for CI, and the records do not measure reviewers' hours or backlog, so continued merging does not show that the review load was sustainable. The classification of the 45 standing corrections was model-assisted, with no second coder.

\textbf{Scope and my role.} This is one startup, one repository, and three people, on GitHub; the identity problem applies to any platform where agents use people's credentials, but the specific fields and protection rules are GitHub's. I directed the agents, held administrator rights, chose which incidents to record, and wrote this paper with AI assistance; using saved records reduces reliance on my memory but does not remove selection bias, and an engineer re-running the gate script would be a stronger check. The engineers' answers come from a team in which anonymity is limited and I am their manager, which may temper criticism. No auditor has assessed the evidence.

%% file: sections/conclusion.tex
\section{Conclusion}\label{sec:conclusion}
Coding agents worked through the same accounts that represented people in our repository, and in the gated phase a session submitted every approval my account gave on a colleague's pull request. The platform recorded those approvals exactly as it would have recorded mine. Recomputing the gate confirmed that protection held where it applied, but on 71 merges into \texttt{main} the approval it checked had been submitted by an agent, and outside the protected branches six merges went ahead on stale approvals before their checks had run. We recorded no one's decisions, and they cannot be reconstructed now; Tables~\ref{tab:auditor} and~\ref{tab:checklist} list what a team should record instead.

%% file: sections/backmatter.tex
\section*{Data Availability}
The study uses proprietary repository records and local agent transcripts. Client confidentiality and commercial constraints prevent release of the product source and the transcripts. We will release the aggregate tables and the gate-analysis scripts, which read only the GitHub REST API, with an archived identifier in the final version.

\section*{Acknowledgment}
I thank Andrea Gualtieri, reader and editor, and the two engineers who answered the survey.

\textbf{Use of AI tools.} Coding agents are the subject of this study. Under my direction, Claude (Anthropic) helped with the data analysis and with drafting and editing the text of all sections, and the classification of standing corrections was model-assisted (Section~\ref{sec:method}). I checked every reported figure against the study data and am responsible for the manuscript and its claims.